\documentclass[12pt,preprint]{emulateapj}

\def\kms{~km~s$^{-1}$\ }
\def\kmsc{~km~s$^{-1}$}

\def\arcs{\char'175\ }

\def\hub{\ifmmode H_\circ\else H$_\circ$\fi}

\shorttitle{Highly Negative Velocity Star in M31}
\shortauthors{Caldwell et al.}

\begin{document}

\title{A Star in  the M31 Giant Stream: the Highest Negative Stellar Velocity Known }

\author{Nelson Caldwell} 
\affil{Center for Astrophysics, 60 Garden Street, Cambridge, MA 02138, USA
\\ electronic mail: caldwell@cfa.harvard.edu}

\author{Heather Morrison}
\affil{Department of Astronomy,
Case Western Reserve University, Cleveland, OH 44106-7215
\\ electronic mail: heather@vegemite.case.edu}

\author{Scott J. Kenyon} 
\affil{Center for Astrophysics, 60 Garden Street, Cambridge, MA 02138, USA
\\ electronic mail: kenyon@cfa.harvard.edu}

\author{Ricardo Schiavon}
\affil{Gemini Observatory, 670 N. A'ohoku Place,  Hilo, HI 96720, USA \\ electronic mail:
rschiavo@gemini.edu}

\author{Paul Harding}
\affil{Department of Astronomy,
Case Western Reserve University, Cleveland, OH 44106-7215
\\ electronic mail: paul.harding@case.edu}

\author{James A. Rose}
\affil{Department of Physics and Astronomy, University of North Carolina,
  Chapel Hill, NC 27599, USA \\ electronic mail:
 jim@physics.unc.edu}

% Use \authorrunning{Short Title} for an abbreviated version of
% your contribution title if the original one is too long

\begin{abstract}
We report on a single star, B030D,  observed as part of a large survey of objects in
M31, which has the unusual radial velocity of  $-780$ \kmsc.  Based on details
of its spectrum, we find that the star is an F supergiant, with a circumstellar
shell.  The evolutionary status of the star could be one of 
a post-mainsequence close binary, a symbiotic nova, or less likely, a
post-AGB star, which
additional observations could help sort out.  Membership of the star
in the Andromeda Giant Stream can explain its highly negative velocity.

\end{abstract}


\section{Introduction}

In large surveys, often a few objects will be found that are unusual
in one respect or another. Our spectroscopic survey of candidate star
clusters in the field of M31 \citep{PaperI} has provided such a
case. From the 1400 objects observed, we expected (and found) a large
number of misclassified stars, due to the heterogeneous quality of the
imaging material used to identify cluster candidates. 

All but a few of these stars are in the foreground, as revealed by
their dwarf spectral types.  Our catalog magnitude limit of V=20
required that any M31 stars observed would have to be bright giants or
supergiants.  Of the few stars we observed in that area that were
clearly not from the foreground Milky Way disk, one star proved to be
unusual in two ways: its spectrum is that of an F supergiant with weak
Balmer emission, and its heliocentric radial velocity is $-780$ \kmsc, which is
the most negative velocity yet found for a single star.  The star is
called B030D in the Bologna catalog of M31 objects \citep{galleti2} , and 
``Bol D30'' in SIMBAD.

This star is unlikely to be a member of M31's disk: it lies close to
the center of M31 on the SW side, at a place where the heliocentric velocity of
the disk is $-425 $ \kms (the minimum disk velocity for M31
reaches $-550$ \kms).  A difference of even 230 \kms 
with respect to the
disk circular velocity is much too high for any disk or thick disk
star in M31. While M31's halo velocity distribution is not well known,
a velocity of $-480$ \kms with respect to M31's center (which has V= $-300$ \kms)
is very high, even for
a halo star.

Could it be an unusual star in the halo of the Milky Way?  M31 lies at
$l$ = 121.2, $b$ = $-21.6$. Accounting for the Sun's motion with
respect to the LSR and the LSR's motion around the galactic
center\footnote{We have assumed that the Sun's motion with respect to
  the LSR is (U,V,W) = ($-9$, 12, 7) \kms and the LSR velocity is 220
  \kms \citep{mihalas}} reduces the velocity to $-$614 \kms with respect to the Milky
Way center.  A recent estimate of the Galaxy's escape speed \citep[544
  \kmsc,][]{martinsmith} suggests that it is unlikely that this star is
bound to the Milky Way; and its highly {\it negative} velocity and
galactic longitude imply that it cannot be a hypervelocity star
ejected from the galactic center such as those identified by
\citet{wbrown}.

This paper describes the analysis of the unusual spectrum of the star
B030D and the reasoning that led us to conclude that the star is most
likely a member of the M31 halo's ``giant stream'' \citep{ibata01} 
close to the pericenter of its orbit.

\section{Spectroscopy}

\begin{figure*}[ht]
\includegraphics[scale=1.0,clip=true]{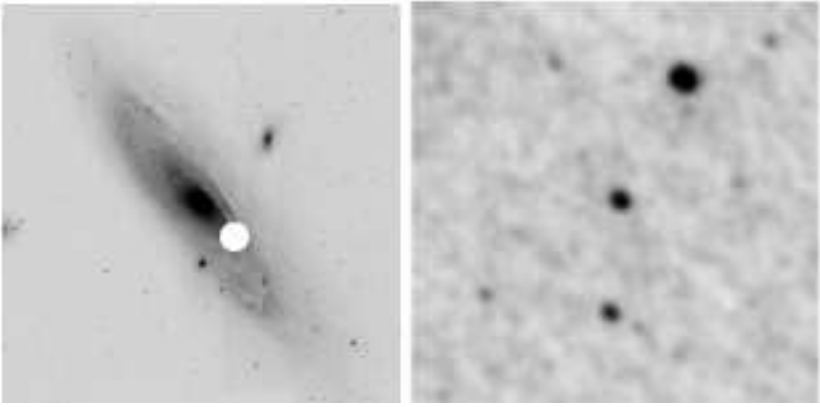}
\caption{Location of B030D in M31. The left image comes from a mosaic
made from J band Digitized Sky Survey images, and covers 2.5\degr . The right image
comes from a V band image of the Local Group Survey of \cite{massey}, and the field size is 30\arcs .
North is up, and east is to the left.
 \label{m31}}
\end{figure*}

B030D resides about 0.3\degr \ SW of the center of M31 (4 kpc in
projection, see Fig \ref{m31}), and was originally 
cataloged by \cite{battistini80} as a ``bluish and diffuse'' object,
based on photographic plates.
It was first observed spectroscopically by us as part of our
MMT Hectospec spectroscopic study of star clusters in M31 \citep{PaperI}. 
From this
good (but not ideal) spectrum we immediately realized that the object
was not a star cluster, which is confirmed by the images made 
available as part of the Local Group Galaxy Survey \citep[LGGS,][see our Fig. \ref{m31}]{massey}.
We also
learned of its unusual velocity and
estimated its spectral type to be F1, though its luminosity class was
difficult to discern.  The higher order Balmer lines were narrow, as would be the
case for a low gravity star, but the H$\beta$ absorption line was
weaker than expected for such a star, and H$\alpha$ was very weak.

\begin{figure*}[ht]
\includegraphics[scale=0.4,clip=true]{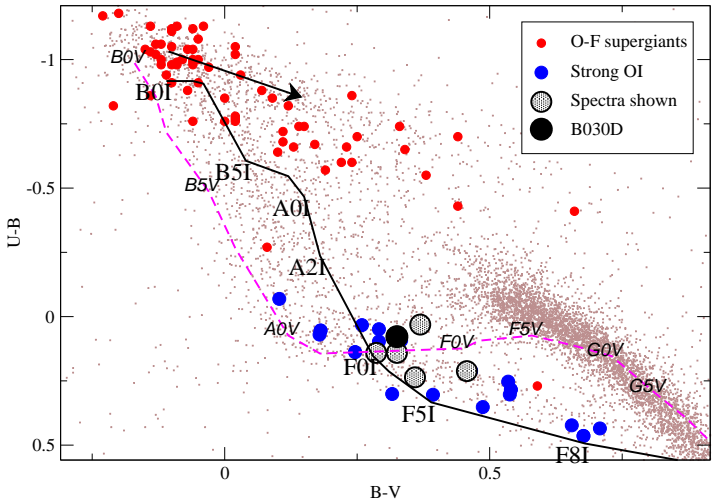}
\caption{UBV colors of stars with V$<19.5$ in the M31 field, 
photometry from \cite{massey}. No reddening corrections have 
been applied. Stars observed spectroscopically to be members 
of M31 are highlighted.
Of those members, the ones whose spectra
are shown in 
Fig \ref{members_blue} and succeeding figures are further indicated with larger symbols.
B030D is shown as a large filled circle.  The sequence of dwarf stars (class V) and
supergiants (Ia) are shown as dashed and solid lines, respectively, the data coming
from \cite{fitz} to which a reddening of E(B$-$V)=0.13 has been 
applied.  The black
arrow indicates the effect of a reddening of  E(B$-$V)=0.25.
 \label{ubv}}
\end{figure*}

We used the UBV photometry of \citet{massey} to obtain more
information about the nature of the star. Table \ref{ubvtable}
summarizes UBV photometry and other basic information on this star.
Here we see that B030D has V=19.02. Figure \ref{ubv} plots U$-$B vs
B$-$V for all stars from the M31 catalog of \citet{massey} with V
brighter than 19.5. B030D is shown by a large black circle.

Where do foreground stars from the Milky Way lie in this diagram? In
this magnitude range, and at the relatively low galactic latitude of
M31, we would expect to see stars from the thick disk and the halo
of the Milky Way.
The dashed line in Figure \ref{ubv} shows the sequence of solar
abundance main sequence stars from \cite{fitz}, reddened by the likely
foreground reddening to M31, E(B--V)=0.13 \citep{massey}.  Metal poor
dwarfs will be found above the dashed line, and indeed the sequence of
Milky Way halo dwarfs can be seen above the line, terminating near
B$-$V=0.5, U$-$B=--0.15.  B030D's B$-$V is too blue to belong to the
old halo of the Milky Way unless it is a field star blue straggler
\citep[e.g.][]{gwp}. Blue stragglers are metal-poor halo stars which are
either significantly younger than most of the halo or the product of a
merged binary.  Judging from this diagram, B030D's UBV colors
are at best merely consistent with it being a main sequence star.

However, there are other options. Where do young supergiant stars from
M31 appear in this diagram? Their position depends strongly on their
reddening. Massey's spectroscopically confirmed M31 O$-$F supergiants
are shown in Figure \ref{ubv} as red solid points, and the supergiant
locus from \citet{fitz} is shown as a solid black line. This locus has
also been reddened by 
E(B$-$V)=0.13. The reddening line, plotted above the
supergiants, shows where more highly reddened supergiants would appear
in this diagram. It can be seen that most of the supergiants from
Massey et al are highly reddened early type stars. This is not
unexpected: we would expect such young stars to be in dusty star
forming regions and so be reddened by dust in both the Milky Way and
M31.   However more recently, 
\cite{drout09} have studied candidate yellow supergiants in M31, and
have identified a number of stars with strong absorption from 
the triplet of O I$\lambda7774$, a sensitive indicator of low
surface gravity \citep{osmer,  arellano}.  These stars are certainly
members of M31, and are shown in Figure \ref{ubv} as blue circles. Several
of these stars have similar colors to B030D.

Unfortunately, B030D lies in the part of
the U$-$B vs B$-$V diagram where the main sequence and the supergiant
lines cross.  
Thus we see that the UBV colors of B030D are consistent with either
a very low gravity star such as an M31 supergiant, albeit one with very little
reddening from M31's disk, or with a halo blue straggler from the
Milky Way.  (We have already noted, however, that its velocity
is so high that if it were in the Milky Way it would be unlikely to be
bound to our Galaxy.)

We subsequently observed the star with Hectospec in November 2007,
along with a number of M31 stellar targets as comparisons, selected on
the basis of color and magnitude to bracket the properties of B030D.

For the second set of observations  we again used 
the Hectospec multi-fiber spectrograph on the MMT \citep{fab}, with
exposure time  4800s on target and 2400s off target to
supply a local background to subtract from the spectra. The spectra
were reduced in the standard way for fiber spectra \citep[e.g.,][]{PaperI}, except that
those off-target exposures were used as local backgrounds. The spectra
have a resolution of 5.0\AA \ and cover the region 3700-9200\AA.
We note that in regions where the bulge of M31 is prominent, such as is the case
for B030D, not subtracting a local background can affect the derived
velocity by as much as 100 \kmsc.  Velocities were found using template
spectra made from the M31 observations described in \cite{PaperI}.  The
spectra were flux-calibrated using spectra of flux standard stars, but
for our comparison purposes here, we rectified the spectra using
arbitrary but consistent continuum settings.

These comparison spectra help sort out the uncertainty in UBV diagram
discussed
above. Stars in the ``main sequence'' locus that starts at bottom
right and reaches to the center of the UBV plot show the broad Balmer
lines and weak gravity-sensitive lines typical of dwarf stars. Their
velocities are consistent with membership of the thick disk and halo
of the Milky Way, with the redder (closer) stars having thick disk
kinematics and the bluer (more distant) stars having halo kinematics. 
Stars in the ``supergiant'' locus show narrow
strong lines, and the presence
of gravity sensitive lines such as Ti II, Fe II$\lambda4172$, Sr II
$\lambda4077$ and the  O I$\lambda7774$ triplet.  
The velocities of those stars are consistent with membership of the
M31 disk.

\begin{figure*}[ht]
\includegraphics[scale=0.4,clip=true]{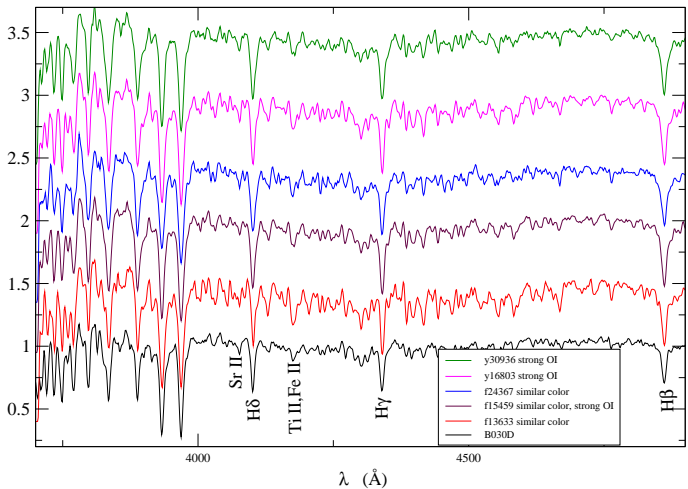}
\caption{Blue spectra of M31 stars, membership being determined by velocity,
Here we show spectra with Ca II ratios and/or color similar to B030D. Note
the similar narrow widths of the Balmer lines, but also the 
weak H$\beta$ in B030D compared to  its
own other Balmer lines.  The spectra in this figure and subsequent figures have
had the continua removed for ease of display. \label{members_blue}}
\end{figure*}

\begin{figure*}[ht]
\includegraphics[scale=0.4,clip=true]{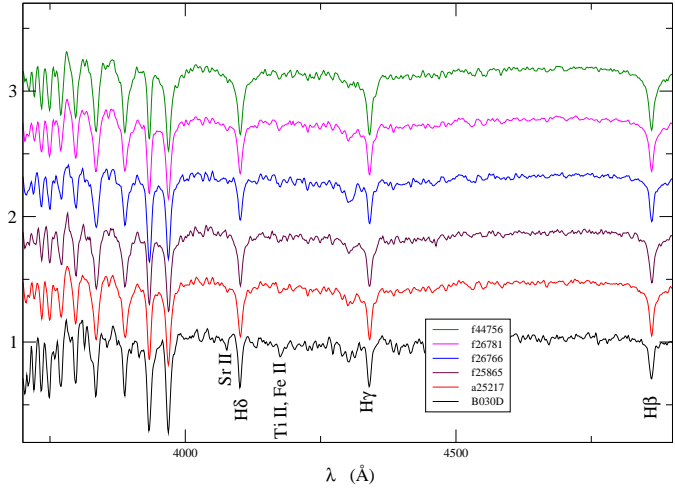}
\caption{Foreground galactic stars, also determined by velocity, with  Ca II ratios and
colors similar to B030D. Note the much broader Balmer lines in the foreground stars. \label{nonmembers_blue}}
\end{figure*}

We selected a few spectra for detailed comparison here, based on the similarity of
their UBV colors to B030D's.  These stars are shown in Figure \ref{ubv} with
large symbols.  Figures \ref{members_blue} and \ref{nonmembers_blue}
show the blue spectra of B030D compared to these M31 supergiants and
Milky Way dwarfs, respectively.  It can be seen that the Balmer lines
of B030D are as narrow as those found in the M31 supergiants,
distinguishably narrower than the foreground dwarfs. Also, the
Ti II, Fe II$\lambda4172$ and Sr II$\lambda4077$ lines are readily apparent in
B030D and difficult to see in the foreground dwarfs.  Finally, the CH features
contained in the G band are partially resolved in B030D; where the G band is
visible in the foreground dwarfs, it is not resolved. All of these
spectroscopic indicators suggest a low gravity for B030D.
Note, however,
the relatively weak H$\beta$ line in B030D, compared to the other M31
stars.

\begin{figure*}[ht]
\includegraphics[scale=0.4,clip=true]{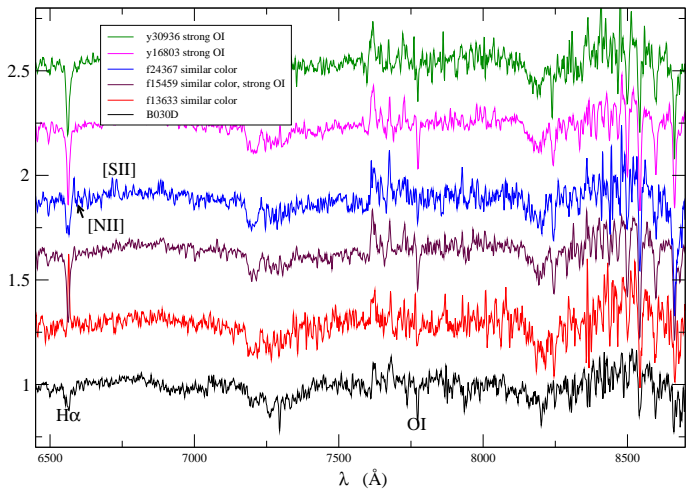}
\caption{Same as Fig. \ref{members_blue} in the red. Note the  strong O I line in all these, indicative
of very low surface gravity.  Also note the filled-in H$\alpha$ line of B030D, and the H$\alpha$ emission for one other star plotted right above it. Neither star shows forbidden emission lines that would indicate
a local HII region, such as is found in the third spectrum from the top, which reveals [NII] emission
as well as  H$\alpha$. \label{members_red}}
\end{figure*}

\begin{figure*}[ht]
\includegraphics[scale=0.4,clip=true]{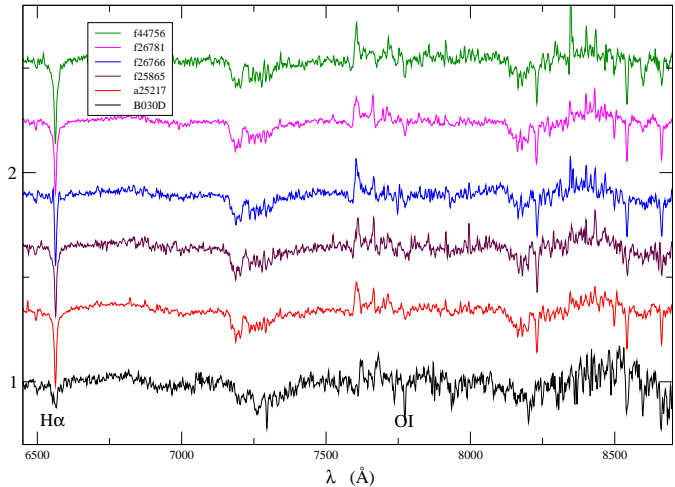}
\caption{ Same as Fig. \ref{nonmembers_blue} in the red. B030D has much stronger O I than the foreground stars. \label{nonmembers_red}}
\end{figure*}

\begin{figure*}[ht]
\includegraphics[scale=0.5,clip=true]{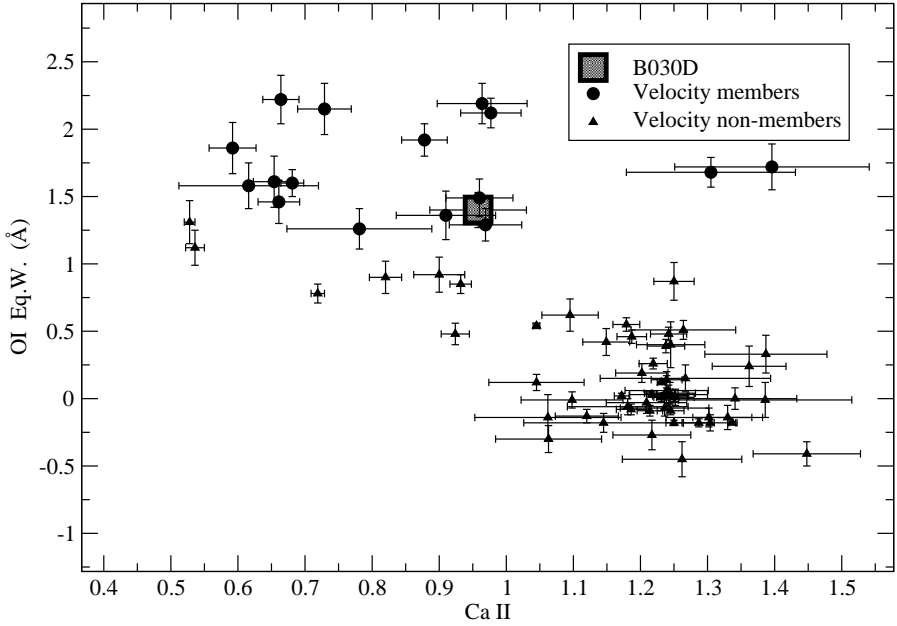}
\caption{Equivalent width of the O I$\lambda7774$ line vs the ratio of the Ca II H\&K lines. The former
is sensitive to surface gravity and the latter
is a useful temperature index. \label{OI}}
\end{figure*}

Turning to the red parts of the spectrum (Figures \ref{members_red} and \ref{nonmembers_red}), 
again we see that the
O~I$\lambda7774$ line is easily visible in B030D and the M31
supergiants, and absent or weak in the foreground dwarfs. Figure
\ref{OI} shows the results of measuring the equivalent width of that
line for all of our high S/N ratio stars, plotted against the ratio of
the Ca II H\&K line strengths \citep[as a proxy for spectral type - the index
saturates for spectral types later than G,][]{rose}.  
G and K foreground dwarfs form the
group around Ca II = 1.25 and O I = 0.  The M31 F supergiants, along
with B030D, clearly separate from the galactic mainsequence stars.  In
view of our qualitative approach here, we do not try to derive an
accurate luminosity for B030D using the relationship between M$_{\rm
  V}$ and O I$\lambda7774$ equivalent width derived in \cite{arellano},
but we do note that their relation and our O I measurement gives an
M$_{\rm V}$ consistent with $-5.5$ (see below).

In summary, the spectroscopy confirms that B030D is a high luminosity
star, at the distance of M31, not a main sequence star from the Milky
Way halo. However, its lower order Balmer lines are unusually weak.

\section{An F supergiant with a Shell}

\begin{figure*}[ht]
\includegraphics[scale=0.5,clip=true]{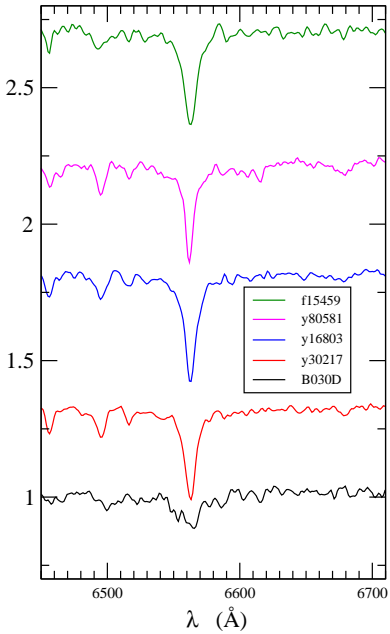}
\caption{H$\alpha$ spectrum, showing weak emission in B030D compared to other M31 member stars which have similar color and luminosity \label{halpha}}
\end{figure*}

Though we have identified the star as an F supergiant, there remains 
the detail of the weak
Balmer absorption.  Figure \ref{halpha} is an
enlargement of the H$\alpha$ part of the spectrum of B030D and a few
comparison M31 F supergiants.  Clearly there is blue-shifted Balmer
emission at H$\alpha$ in B030D, and that also explains why the H$\beta$
line is weaker than expected. Since there are no accompanying
forbidden lines, and moreover the offset exposures show no emission at
all, we must assume the emission is not from the surrounding area but
from the star itself, and likely
due to a shell. Shell emission is common among the rare, luminous 
high-latitude Galactic F variable stars, such as 89 Her \citep{cli} and 
HR4912 \citep{luck}.

\section{Evolutionary Status of B030D}

The V magnitude of the star is 19.0. Using the distance modulus
of 24.43 \citep{freedman} and reddening of 0.13 \citep{massey}, 
we derive M$_{\rm V}=-5.5$.
B030D thus lies in an interesting evolutionary state. With an effective 
temperature of $\sim$ 6500~K (estimated from its spectral type)
and a luminosity of $\sim$ 15,000 L$_{\sun}$ (log L = 4.2). 
its position in the Hertzsprung gap guarantees that it is evolving 
rapidly through the HR diagram. Here, we consider four likely options 
for its evolutionary status and propose new observations that can 
distinguish among them.

\noindent
{\it Luminous Blue Variable (LBV):}
M31 contains $\sim$ 20 LBV's, evolved massive stars which vary 
erratically in brightness and undergo periods of enhanced mass loss 
\citep[e.g.,][and references therein]{massey07}. Although B030D is 
roughly 10 times fainter than the typical LBV, its optical spectrum
resembles spectra of S Dor in its cool state \citep{massey2000}.
Compared to a `cool' S Dor, the Balmer  emission in B030D is weaker
and the optical colors much redder. 

\noindent
{\it Post-Main-Sequence Close Binary:}
\citet{bath1979} proposed that the physical appearance of an evolved close 
binary can be indistinguishable from an LBV.  In this mechanism, a 
main-sequence secondary star accretes material at a high rate from a 
Roche lobe-filling post-main-sequence primary star. A luminous disk 
surrounding the secondary produces the F-type spectrum and erratic 
brightness variations; mass lost from the disk forms an extended shell 
that produces Balmer and other low excitation emission lines. 
\citet{gallagher1981} noted that this mechanism is a convenient 
explanation for AF And and other LBV's outside the spiral arms of a 
star-forming galaxy. The high luminosity in B030D requires an accretion
rate of $\sim 10^{-3}$ M${_\sun}$ yr$^{-1}$ onto a $1-4$ M$_{\sun} $\ main-sequence-star.
This high accretion rate can produce an F-type optical spectrum with 
normal optical colors \citep[e.g., the symbiotic binary CI Cyg;][]{kenyon91}. 

\noindent
{\it Post-Asymptotic Giant Branch Star:}
Post-AGB stars are high luminosity stars that have recently ejected a 
planetary nebula. As they eject their hydrogen envelopes, they evolve 
at roughly constant luminosity to higher effective temperatures.
Although many post-AGB stars have much hotter temperatures than B030D,
\citet{hrivnak1989} show F-type optical spectra for several post-AGB
stars with luminosities similar to the luminosity of B030D. Nearly all 
post-AGB stars are surrounded by extended shells of dust and gas; thus, 
these stars often have large infrared excesses from the surrounding
dust and radio emission from the gaseous shell.  
With M$_{\rm V}$ = $-$5.5, B030D is brighter than the typical post-AGB star 
\citep{bond, sahin}. For the solar metallicity
typical of M31 \citep{venn}, stars with initial masses of 4 $M_\odot$ produce 
post-AGB stars with white dwarf core masses of 0.78 $M_\odot$ and log L 
= 4.2 \citep{lawlor} . However, the M31 Giant Stream (see below)
probably 
has a lower metallicity of roughly 1/3 solar \citep{raja}. For this metallicity, the 
mass of a post-AGB progenitor with log L = 4.2 is roughly 3--3.5 $M_\odot$.

\noindent
{\it Symbiotic Nova:}
The symbiotic (or very slow) novae are eruptive variables in wide binaries 
with a red giant primary. In these systems, the white dwarf secondary
accretes matter from the red giant and undergoes occasional thermonuclear
runaways. During an optical maximum which can last many decades, a symbiotic 
nova has an F-type supergiant spectrum with Balmer emission lines 
\citep[e.g., PU Vul; ][]{kenyon1986}. 
Although symbiotic novae occupy the same region of the H-R diagram as 
post-AGB stars  \citep{kenyon1983}, they often have much smaller
shells of gas and dust.

Without additional data, choosing the most likely of these possibilities 
is difficult. However, the relatively normal U$-$B colors suggest the 
post-AGB and symbiotic nova interpretations are more likely than the
LBV-like possibilities. Higher resolution optical spectra would provide
a useful diagnostic. LBV's typically show higher velocity and more variable 
winds than post-AGB stars or symbiotic novae. 
Many post-AGB stars and symbiotic stars are spectroscopic binaries with periods 
of 1-3 yr. Measuring an orbital period from radial velocity data could rule out
several alternatives.
Infrared spectroscopy should provide a clear discriminant between a
post-AGB star and a symbiotic nova. In a symbiotic nova, infrared
spectra show strong CO absorption lines characteristic of red giant
photospheres \citep[e.g.,][]{schild}.  In a post-AGB star, CO 
features at 2.3 and 4.6 $\mu$m suggest the much cooler excitation 
temperatures (T $\sim$ 300\degr \ K) expected from a circumstellar shell 
\citep[e.g.,][]{lambert, hinkle}.

There is one similar star we have observed in M31, with H$\alpha$  in emission but
no other nebular lines: f13633 (J004025.26+403926.9), which has the same luminosity but perhaps lower
surface gravity than B030D. It is located in a one of the spiral arms, and appears to
be double, in the LGGS image.

\section{Velocity}

We concluded in Section 2 that B030D is a high luminosity star at
roughly the distance of M31. In Section 4 we went on to discuss four
possible options: a very young, massive star type (a luminous blue
variable) or three stellar types which are associated with older
populations (a post-mainsequence close binary, a post-AGB star or a
symbiotic nova).

Projected close to M31's center but with a 355 \kms velocity
difference with disk stars there (see \ref{ubvtable}), 
it would seem more reasonable to
associate the star with M31's halo. However, studies of the age
distribution of M31's halo show no evidence for very young stars
\citep{tombrown}, so if the star is a luminous blue variable, we need
to seek possible mechanisms to accelerate a young disk star to such a
high velocity. There are a number of such possibilities, such as the
interaction of a binary with the super-massive black hole in the
center of either M31 or M32, which would disrupt the binary and
produce a ``hyper-velocity star'' \citep{wbrown}, interactions in
the core of a dense star cluster, or a ``runaway'' caused by a
supernova explosion in a binary system \citep{leonard}. While such
interactions are likely to be very rare, they are not impossible.

\begin{figure*}[ht]
\includegraphics[scale=0.7,clip=true]{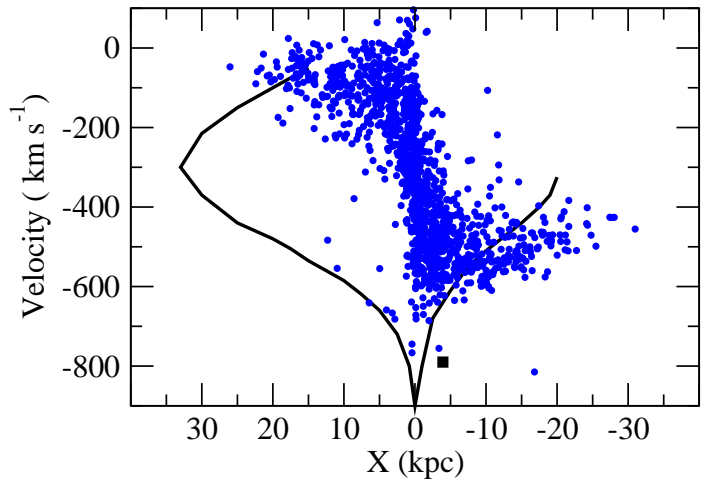}
\caption{Velocities of PNe and B030D (dark square) plotted against distance along the major axis, for objects within 1 kpc of the major axis ($|$Y$|<$ 1 kpc). The orbit for
the M31 Giant stream proposed by \cite{merrett03} is also shown as a solid
line, revealing that
B030D can be considered part of the stream.  \label{stream}}
\end{figure*}

However, there is a simpler, more likely explanation for the star's
large velocity which involves a well-studied component of M31's halo,
the Giant Stream, which was first discovered using star counts by
\citet{ibata01}. Subsequent kinematical studies
\citep[e.g.][]{merrett03,ibata04,merrett06} have shown that the stream
has a highly elliptical orbit with perigalacticon very close to M31's
center (within 2 kpc or less) and apogalacticon at 125 kpc
\citep{mcconnachie03,ibata04,fardal07}. Models fit to the stream have
suggested velocities at its perigalacticon as low as V$_{\rm helio} =
-900$ \kms \citep{fardal07}. The planetary nebulae studied by \citet{merrett06}
include ones that are possible stream members with V$_{\rm helio}$ as low as
$-838$ \kmsc, although this PN has only
one line at 4992\AA \ (from a Hectospec spectrum of ours), thus
the possibility that the object is actually at high redshift cannot be
dismissed.  In Figure
\ref{stream}, we show velocities of PNe, drawn from \cite{merrett06}
and our Hectospec observations (paper in preparation), along with the
orbit of the stream proposed in \citet{merrett03}. Clearly some PNe
fall along the orbit of the stream, as does B030D.

Stars with masses of 3-4 $M_\odot$ have main sequence lifetimes much shorter 
than the youngest stars in the Giant Stream. Thus, a post-AGB member of the
Giant Stream with M$_{\rm V}$ = $-$5.5 is unlikely. However, B030D could be a 
post-AGB runaway star ejected from the disk of M31 at high velocity. In the 
Milky Way, 5\% to 10\% of 3--4 $M_\odot$ B-type stars are runaways (Stone 1991).  
Although runaways with velocities of 200$-$300 km s$^{-1}$ relative to the disk 
are rare \citep{stone, martin}, theoretical simulations produce high velocity 
runaways at the projected distance of B030D from the center of M31 \citep[e.g., the Milky
Way study of][]{bromley2009}.  In this case, B030D would have evolved into a post-AGB star
after ejection as a runaway.

The post-AGB possibility seems remote and we conclude that 
the most likely origin for the star is as part
of the Giant Stream. \citet{tombrown} show that the stream contains
stars as young as 4 Gyr, so the two older options discussed in
Section 3    (a post-mainsequence close binary or a
symbiotic nova) are the  best possibilities.

\section{Summary}
B030D is a curiosity. Not only does it have an extremely negative
velocity, but it is an unusual and rare kind of star as well.  In our
research into spectroscopic databases, we could not find any exact analogues
that matched the details of its spectrum with regard to the Balmer emission.
That problem was certainly mostly due to the lack of modern, digital spectra of the classes
of unusual stars described above, but the short life of this kind of F shell star must also
be part of the story.  In addition to IR observations, a 
further piece of data that may help to determine the
exact classification would be the star's variability.  We examined the star on both epochs of
the Palomar Sky Survey, but could not find a significant change in brightness over
that period (36 years), though the different passbands used for the two surveys makes
a comparison difficult. Modern digital images of M31 could of course provide more 
stringent limits on variability.

\acknowledgments
We would like to thank Dan Fabricant for leading the effort to
design \& build the Hectospec fiber positioner and spectrograph,
Perry Berlind \&  Mike Calkins for help with
the observations, John Roll, Maureen Conroy \&  Bill Joye for their many 
contributions to the Hectospec software development project. We
acknowledge important
discussions with Warren Brown, Margaret Geller  \&  Phil Massey.

HLM was supported by NSF grant AST-0607518.
Work on this project has also been supported by HST grant GO10407.

\clearpage

\clearpage
\pagestyle{empty}
%\setcounter{page}{0}
%\vspace*{1.5in}
%TABLES
\begin{deluxetable}{llllrrllrll}
%\rotate
%\tablenum{1}
\tablecolumns{11}
\tablewidth{0pc}
\tablecaption{Basic Data on B030D and Comparison stars \label{ubvtable}} 

\tablehead{\colhead{Object} &\colhead{Massey(2006) ID} & \colhead{RA}  &\colhead{Dec}  &\colhead{Velocity}  &\colhead{O$-$C\tablenotemark{a}}  &\colhead{V}   &\colhead{B$-$V}  &\colhead{U$-$B} &\colhead{V$-$R} &\colhead{R$-$I} \\
\colhead{} & \colhead{} & \multicolumn{2}{c}{2000} &\colhead{\kms}   &\colhead{\kms}   &\colhead{} &\colhead{} &\colhead{} &\colhead{} &\colhead{}
}
\startdata
M31 members&&&&&&&&&\\
B030D &J004141.49+410308.2 &  0:41:41.5   &  41:03:08.0  & $-780\pm 20$ & $-355$& 19.02 &   0.33 &  0.08 &   0.22 &    0.27 \\   
f15459 & J004030.62+404523.8  &  0:40:30.6   &  40:45:23.8  & $-559\pm 20$ & $-16 $& 18.33  & 0.29 &    0.14  & 0.18  &  0.22 \\
y16803 & J004034.00+405358.3& 0:40:34.0   &  40:53:58.3  & $-505\pm 18$ &$-12 $&  18.02  & 0.46 &    0.21&   0.27 &   0.29 \\
f13633 &J004025.26+403926.9&  0:40:25.3   &  40:39:26.9  & $-640\pm 34$ &$-100 $& 19.16&   0.33&     0.14  & 0.26 &   0.30 \\
f24367 &J004117.90+404845.0  &   0:41:17.9  &   40:48:45.0 & $-510\pm 17$ &$-6 $  &19.10  & 0.37   &  0.03  & 0.25  &  0.29\\
y30936 & J004201.09+403951.9 &  0:42:01.1  &   40:39:51.9 & $-419\pm 17$ & $-13 $& 18.46  & 0.36    & 0.24  & 0.26  &  0.36\\
\hline
Foreground&&&&&&&&&\\
a25217 &J004123.97+405249.0 &  0:41:24.0  &  40:52:48.7 & $-51\pm 18$ &\nodata & 16.73  & 0.42    &-0.05 &  0.27   & 0.22\\
f25865 &J004128.60+410229.6 &  0:41:28.6   &  41:02:29.6 & $-63\pm 18$ & \nodata&  18.29 &  0.43   & -0.10 &  0.28  &  0.37\\
f26766  &J004134.97+410537.0 &  0:41:35.0  &  41:05:37.0 & $ -76\pm 21$ &\nodata & 17.94  & 0.50    &-0.07  & 0.32   & 0.34\\
f26781  &J004135.14+410116.3 & 0:41:35.1  & 41:01:16.3 & $  9\pm 16$ &\nodata & 16.60  & 0.43    &-0.02  & 0.29   & 0.30\\
f44756 &J004252.79+404605.6  & 0:42:52.8  & 40:46:05.6 & $-63\pm 14$ &\nodata & 17.75  & 0.39    & 0.06  & 0.23   & 0.29\\
\enddata
\tablenotetext{a}{Difference between the observed velocity and the computed local velocity of M31's disk.}
\end{deluxetable}

\clearpage

\end{document}